# A Workshop Series for Effective Use of AI in Uncertain Times: Building a Physics Faculty Learning Community

David Perl-Nussbaum and Noah D. Finkelstein

Department of Physics, University of Colorado Boulder

## Abstract

Generative AI tools are being widely taken up by students in their physics courses and beyond, often before instructors and institutions can develop policies and effective approaches for the use of these tools. Building on a framework for change in the era of AI, we developed and implemented a faculty learning community to help a university physics department address these challenges collectively. Over six biweekly sessions, faculty worked through course policies, classroom conversations about AI, AI-integrated coursework tasks, and assessment. Each session shared a common structure: we presented local data and department-sourced materials, tested them in small groups, and discussed them together, emphasizing durable pedagogical approaches over specific tools and platforms, and leading with evidence of students' own AI use. The series produced a shared, evolving repository of resources for faculty to draw on. This workshop provides an adaptable, theoretically informed model for a faculty learning community that departments can build on.

## 1. Introduction

### *1.1 A new era - starting before we're ready*

A variety of new technology-based tools and enabled practices, such as students using generative AI (genAI) to solve traditional physics problems, are showing up in our classrooms, some with our invitation and many without. Yet in general we find ourselves underprepared for effectively using these tools, both because the evidence-base of effective practice is lacking and because we, as educators, have not been prepared to use them. We are asked to respond before we are ready. As a result, institutions are scrambling to provide frameworks and principles for effective use -- among these are calls for the development of faculty learning communities. As the recent MIT report on AI and education states, we need to "build processes, teams and tools for continuous reflection, iteration, and improvement" (MIT, 2026). To address these challenges and opportunities we present a model and associated resources for engaging physics educators at the tertiary level in productive discussions around artificial intelligence (AI)[1]-induced changes that are arriving in our classroom.

We present a case describing the design and implementation of a faculty learning community that realizes a new model and associated design principles of institutional change that is required in the AI-era (Perl-Nussbaum & Finkelstein, 2026). This framework centralizes the presence of AI as an *arrival technology* (Reich & Dukes, 2025) - disrupting our practices before we understand it, as opposed to prior educational change efforts that have focused on *adoption technologies* - which are based on a strong evidence base that educators and institutions choose to employ. Building on a tradition of institutional change projects around adoption technologies in our department of physics (Chasteen et al., 2015; Corbo et al., 2015; Otero et al., 2010), we built a faculty development program that provides an approach for our department to collectively engage in intentional and systematic use of AI. The workshop was born out of local need and interest and serves to: demonstrate the utility of the AI-based educational change framework, provide one productive response to the current circumstances we find ourselves in, and begin to assemble resources that may be of use to individual faculty and departments alike.

[1] By artificial intelligence we refer to dominantly generative AI (LLM)-based tools, but are also inclusive of machine learning (ML), reinforcement learning (RL), vision and audio processing, and natural language processing (NLP) based techniques.

We do not seek to be comprehensive nor evaluative, as we believe it is premature to present such outcomes. Rather, given the demands of the era -- for us to act before we are ready – this working study seeks to contribute to the discussion of possible responses to changes that are spreading rapidly throughout and across higher education.

### *1.2. (Local) Motivation*

A variety of local motivations for this faculty learning series echo the international discussions around these arrival technologies in our educational spaces. First and foremost, our faculty recognized the appearance of AI-approaches in our classrooms, most often without invitation. A survey at our institution revealed 80% of students were using AI weekly, and were seeking guidance from their instructors (Center for Teaching & Learning, 2025). In the absence of guidance, by our faculty or by our institution more broadly, students were using AI approaches anyhow. Faculty noted that current generation AI tools could correctly solve and explain most undergraduate physics problems. All written and coded based project work was susceptible to AI influence (orienting, assisting, or even replacing student-generated work). Increased concerns were being raised about cognitive offloading (Gerlich, 2025), and outright cheating (Skogvoll & Odden, 2026). A common default response has been to focus more on increased in-class, supervised testing. However, the same historical concerns around testing – such as the authenticity of the practice, the high-stakes nature of such exams, biases, and their inaccuracies as measures of learning remained (Schinske & Tanner, 2014).). Reverting to former practices did not seem to advance our educational approaches. Moreover, over the course of the faculty working group series, the first instance of a student being caught using smart glasses during a departmental exam was reported, paralleling the rise of such tech tools in high stakes testing environments (Corbin et al., 2026).

All the while, our faculty, who are generally limited in their formal preparation to teach, are not sufficiently prepared to effectively respond to, let alone lead, in the effective use of AI-based educational practices. The medium itself is new – many faculty are not familiar with how AI-tools operate or how they change the traditional educational practices -- whether course preparation, class practices, or evaluation practices. Furthermore, with the rapid evolution of these tools, how educational practices change is a moving target. To these ends, our faculty asked for a departmental response.

One semester (Fall 2025) began as an informal discussion series, meeting at least monthly and surfacing faculty concerns and opportunities around these new tools. The main outcome of the discussion series was a request to have a more formal learning series to explore what we might do systematically. This learning series began in January of 2026, and we report its design and produced materials in this paper.

## 2. Designing the workshop: Building on a Framework for Change in the AI era

GenAI tools and practices have entered our classrooms not through slow and intended adoption of evaluated innovations, but through sudden and widespread use – largely before institutions could form any response (Alba et al., 2026). Reich and Dukes (2025) call such technologies *arrival* technologies: they disrupt our practices and bypass our evaluation and dissemination processes. This sets genAI tools apart from the *adoption technologies* behind prior educational reforms, such as Peer Instruction (Mazur, 1997) or Learning Assistants (Otero et al., 2010) – which are based on a strong evidence base that educators and institutions choose to employ. While the medium and resources may be changing, our goals to support equitable, effective and inclusive student learning and engagement have not. In order to support our goals amidst new disruptive technologies, we need to adapt our educational change practices. A promising approach is to empower educators through a faculty learning community while assembling resources and practice that hold the promise to be effective (Cruz et al., 2025; Henderson et al., 2011).

Our workshop design follows our framework for leading institutional change in the AI era (developed in Perl-Nussbaum & Finkelstein, 2026). We argue that the *arrival technology* features of AI have implications for how educational change should be led - both with respect to the *tools* and the *people* involved (Table 1).

With respect to the tools - since AI tools lack the evidence base, are constantly changing, and designed for general use rather than for educational purposes, we argue that change initiatives should: proceed through local and humble exploration (rather than the dissemination of unevidenced practices); focus on durable pedagogical approaches rather than particular platforms and tools; and take up the broader questions that general-purpose technologies raise, from ethics to use beyond the classroom.

With respect to the people - since faculty did not choose AI's arrival and are affected unevenly, since none of the change agents leading AI-reform yet hold settled answers, and since students are already leading in use, we argue that change initiatives should map the various faculty needs and account for their uneven exposure, reposition change agents as facilitators of collective inquiry rather than sources of evidence-based practices, and treat students as partners in change, building on their use and needs.

| Characteristics of AI-driven change | Design implications |
|---|---|
| ***The tools*** | |
| **Evidence base:** AI tools are taken up and scaled before a pedagogical evidence base can form. | ● Humble and local inquiries |
| **Rate of change:** tools are rapidly changing and may be obsolete before dissemination. | ● Approach over tools<br>● Evolving guidelines of use |
| **Scope and intended use:** tools are general-purpose, entering education from the outside. | ● Broadening learning goals<br>● Building flexible and general practices |
| ***The People*** | |
| **Faculty agency:** disruption arrives whether or not faculty opt in – through institution policy or widespread student use. | ● Map disruption and needs<br>● Invest in support structures |
| **Change agent role:** lack of AI educational experts. change agents must respond without established practices to broker. | ● Facilitating collective inquiry<br>● Prioritize pedagogical expertise |
| **Student role:** students lead in use. They are already engaged and seeking guidance from faculty. | ● Students as partners<br>● Build on student use<br>● Equitable and ethical use of AI |

*Table 1: Six dimensions along which AI-driven institutional change differs from prior education reforms, and the design implications that follow (see Perl-Nussbaum & Finkelstein, 2026, for the full framework).*

## 3. Workshop Design

The workshop series ran in the University of Colorado Boulder Physics Department, a large public R1 department with a long history of systematic education reform. It comprised six bi-weekly sessions, 60-75 minutes each, with about ten faculty attending any given session and nineteen participating across the series. Our goals were to address the arrival of AI *as a department*, supporting both faculty and students, and begin to establish shared departmental resources and policies. Building on models of Faculty Learning Communities (Cox, 2004) and Departmental Action Teams (Reinholz et al., 2019), and following our theoretical stance (Perl-Nussbaum & Finkelstein, 2026), the meetings were framed as collective, local inquiry rather than a general faculty training workshop, emphasizing pedagogical approaches over specific tools, and leading with student use. The sessions followed a common structure: We started by sharing local data, whether around AI policies and use or activities gathered from different courses, and then had faculty test and reflect on the materials in small groups, followed by collective discussion.

We describe each of our Spring 2026 faculty sessions: the session structure, design principles, and what surfaced in the meeting – including materials that were curated or created for each session. We do not intend to share these as a validated curriculum. Rather, we frame this as one possible instantiation of a thoughtful (i.e., guided by theoretical framing) initiative to support faculty in response to AI. Our aim is to offer faculty and departments interested in change a useful starting point for initiating AI-related faculty learning communities. Table 2 provides an overview of the sessions, which are then described one by one.

|  | Session | Resources/outcomes |
| --- | --- | --- |
| 1 | Policy statements | Sample policies, categorized by themes/approaches |
| 2 | Discussing AI policy with students | Student survey tool & data, discussion guidelines & clicker questions to engage students |
| 3 | Homework corrections | Samples & guidelines for checking homework with/without AI |
| 4 | Using AI in course tasks and HW | Sample activities integrating AI in coursework (generating code, diagrams, sims, explanations) |
| 5 | AI and assessment | Oral exam structures |
| 6 | Reflection & planning forward | Repository of resources |

*Table 2: Workshop sessions and produced resources.*

### *3.1 Session 1: Course Policies*

Lacking university or departmental policies on AI use, many of our faculty were uncertain about what their policy should be, where to draw their line on appropriate use, and what was realistic to expect from students. As a result, some of our faculty did not give students any guidelines on AI use in their courses[2]. This session provided opportunities to compare possible policies and help instructors reflect on their own approach towards AI use in their course and how it should be framed for the students.

**What we did:** We began by presenting and discussing the campus decision to refrain from any policy statements, leaving individual instructors to come up with their course-specific policies. We then shared a printed set of AI syllabus statements collected from within our department and organized along a spectrum spanning from *no allowed AI use*, to *limited and conditional use*, to *full use.* These policy approaches were contextualized by course topic and audience level. Reading across these examples in small groups, faculty identified common themes, reflected on their own policies (or lack thereof), and considered where they would want to adjust (for example, adding a course policy, or clearer guidelines for disclosing AI-assisted work).

**What surfaced:** We identified shared themes across AI-policies, including framing the course learning goals, specifying allowed and not allowed uses, stating consequences for violation and more. Some of the policy statements explicitly called for students' input to reshape the course policy, or acknowledged the policy would be revisited as tools evolve. The categorized statements and shared themes became the first entries in a departmental repository of resources, to help faculty shape their AI syllabus statements. Participants also agreed that the department should offer a set of syllabus statements templated across the spectrum (from *no use* to *full use*), creating a shared language across courses and consistency for students and faculty, while leaving the pedagogical choice to individual instructors. These were also created and shared through the departmental repository.

---

[2] This was not an uncommon occurrence as of the spring 2026 term (several years into widespread availability of genAI tools).

**Design principles:** Presenting a spectrum of possible policies from our department reflects the *humble and local inquiries* we advocate for. As facilitators, our roles shifted from presenting best practices to surfacing local resources and facilitating collaborative inquiry into them.

*3.2 Session 2: Discussing AI use and policy with students*

This session addressed how to move from a static policy statement to socializing the policy with students and develop an ongoing classroom conversation around AI use and norms.

**What we did:** We began by sharing findings on students' AI use and perceptions from a campus-wide survey and a second, physics-specific survey we issued in one of our large-enrollment classes. Results of the surveys showed that students were asking for instructor guidance on AI use within their disciplinary courses, and that students shared the faculty's concerns about AI circumventing their own learning. Furthermore, most students reported they were using AI not for malicious uses, but rather to help them when stuck (see Figure 1 for examples). These findings shifted the framing of AI policies from an adversarial one - faculty policing student cheating - into a shared problem where both faculty and students navigate AI use towards mutual learning goals. That said, a non-negligible fraction of students engaged in behaviors that many of the faculty found inappropriate (e.g., 29% of students reported sometimes or often using AI to produce an entire solution). Such data were also helpful for seeding faculty discussion with students.

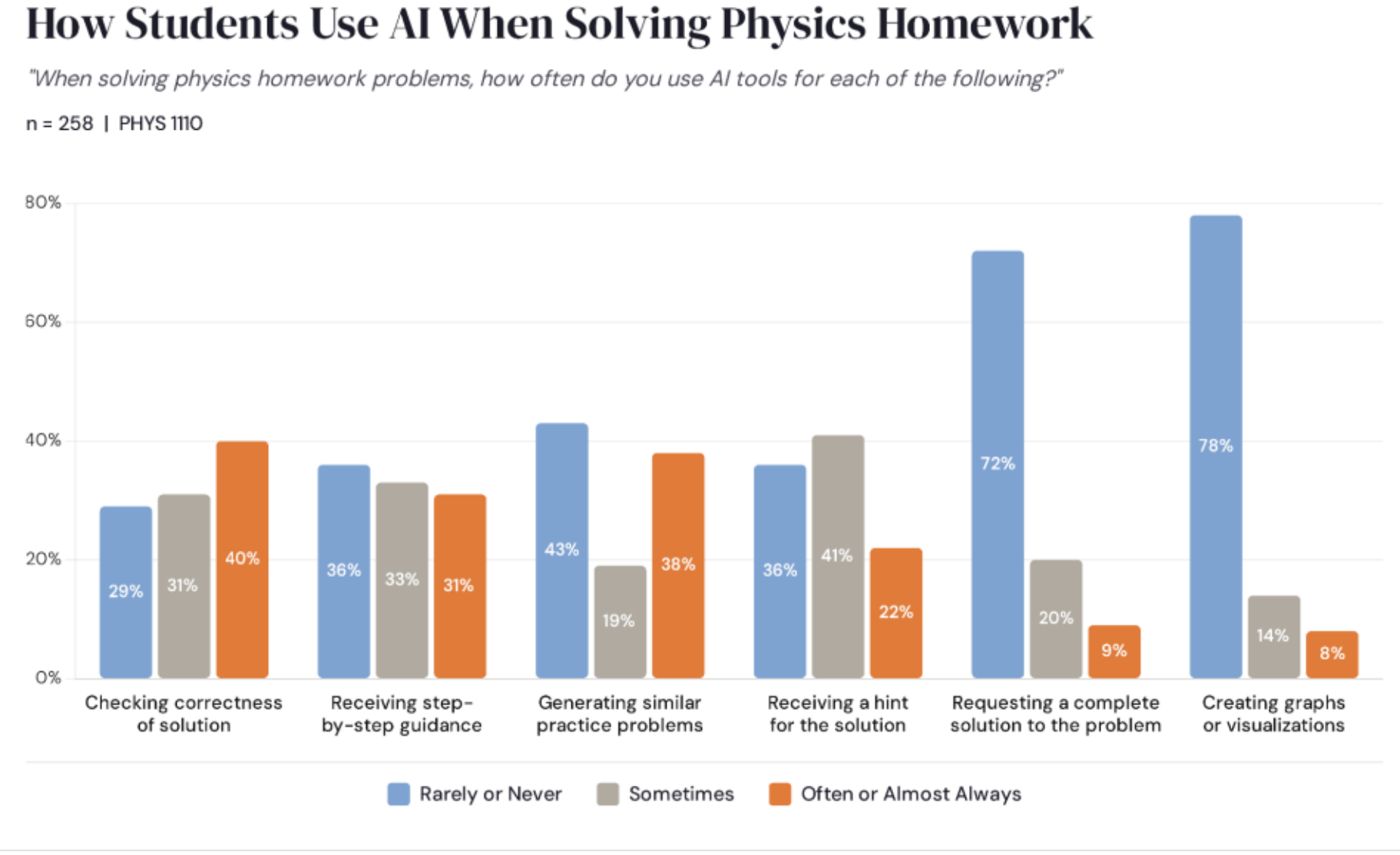

*Figure 1: Sample findings from a physics student survey on AI use and perceptions, administered in PHYS 1110 (General Physics 1, freshman level). Most students report using AI for supportive tasks (e.g., checking correctness or generating practice problems) rather than to complete graded work.*

After discussing the survey results, we shared materials from different courses that engage students in discussion around AI use in general and in the specific course. These materials included the student survey for AI use in physics course work, clicker questions for in-class discussions (e.g. from a computational physics course asking students if they are interested in learning how to integrate AI use to their coding, see Figure 2), and a homework task where students go through the course AI-syllabus statement and comment on it. Faculty reflected on these materials in small groups, shared if and how they were currently engaging students (on a scale of 0: having a policy; 1: presenting it in class, 2: presenting and asking for questions; 3: asking for feedback and engaging in discussions), and identified what additional support they would need to further engage in class-based discussions.

**Class Discussion on AI - Phys 2600**

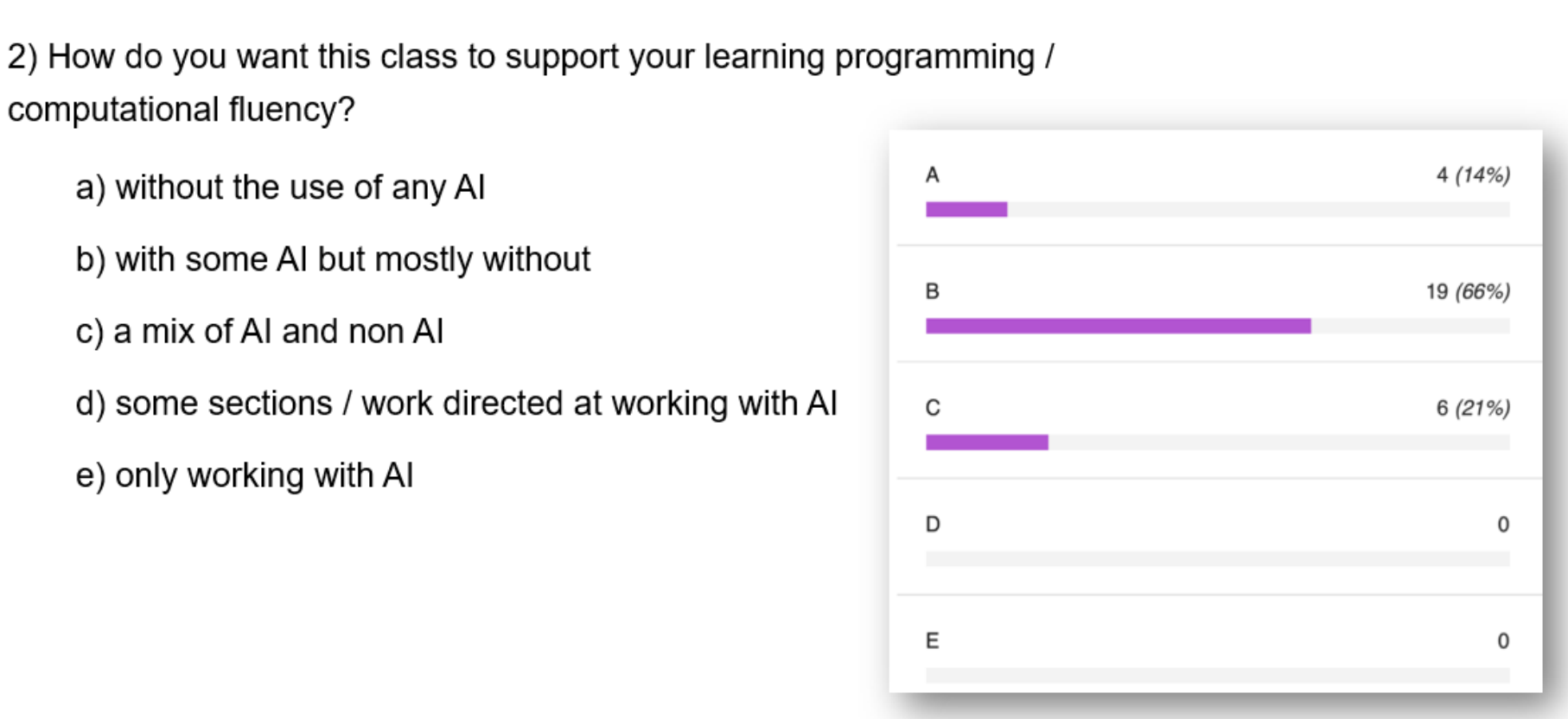

*Figure 2: Example of a clicker question to engage students in a discussion about the level of AI integration they envision in their computational physics course (PHYS 2600, Introduction to Scientific Computing, sophomore level).*

**Design principles:** This session reflects *building on student use.* We were interested in bringing students' voices into consideration, building and responding to their existing AI practices, perceptions and needs.

**What surfaced:** Faculty were interested in running the student surveys in their own courses, both for better understanding students' perspective and as a starting point for dialogue. We ultimately issued the survey in 5 different courses, ranging from middle division to upper division, reaching N~350 students and generating useful insights that were shared continuously in our faculty sessions and with the department more broadly. Similar to session 1, this session also generated a repository entry where we shared the student survey, results from different courses, and several approaches and materials we collected to engage students in discussions around AI norms.

### *3.3 Session 3: Homework corrections with and without AI*

Homework (HW) is one of the course components most affected by AI. Students in our local surveys were reporting that "checking correctness" of their homework solutions is among their most common uses of AI, and that they generally do not view it as a violation of academic integrity. This session moved from policy to a specific formative practice: homework corrections, in which students revisit incorrect problems, identify their errors, and rework them. We asked what role, if any, should AI play in this process, where it supports the intended learning rather than replacing it, and what guidance should we give students around this practice.

**What we did:** We opened with the data we collected showing significant amounts of students' use of AI to check if their solutions are correct before submitting (for example, among undergraduate students in introductory physics, 71% report they do it sometimes, often, or almost always, see Figure 1). We then presented research from our department that compared different approaches and structures faculty had (pre-AI) to encourage students' reflection on their HW solutions (e.g. adding comments on their original submissions after instructor solutions are published and before grading), and showing that students in these courses struggle to accurately identify where they went wrong, even with a full solution (Griston & Wilcox, 2025).

Finally, in small groups, faculty tested AI as a correctness-checker on real (anonymized) student homework answers from a modern physics course (both technical – calculating penetration depth, and conceptual – explaining limitations of the Bohr model). Faculty tried different prompts to see which, if any, encouraged what they perceived as productive reflection practices. Groups discussed the variations in goals for student use of AI in HW corrections – avoid using it, note its limits, test its capabilities, or use it regularly for reflection – and drafted do's and don'ts for students.

**Design principles:** This session continued *building on students' use*: it was designed in response to a frequent use of AI, and that was left unaddressed by faculty. We also emphasized *pedagogical approach over tools*: instead of presenting one of the several existing AI tutoring platforms designed to scaffold student problem solving and reflection (e.g., Dange et al., 2026), we anchored the session on the disciplinary practice itself: reflecting on the correctness of a physics solution (including, for example, checking units or extreme cases), and supported this discussion with education research pre-AI.

**What surfaced:** Faculty's testing of AI chatbots on real student work showed several limits. AI easily solved every problem, but its ability to correct student work and give useful feedback was uneven and depended heavily on prompting (as expected with general-use chatbots); it tended to focus on surface features rather than deeper conceptual issues and needed guidance specifying instructor/question goals and rubrics to be reliable. The group concluded that the priority is developing students' correction and self-checking skills, and that students will need instructor's modeling of productive strategies that can be used to prompt these with AI.

### *3.4 Session 4: Using AI in HW tasks: generating and revising content*

While session 3 examined AI practices for checking students' finished work, this session turned to using AI to produce work by generating code, diagrams, explanations, or simulations (Ben-Zion et al., 2026). At the time of this session, general-purpose tools could do these things well, so the question at hand was not whether AI can generate these resources, but what students should do with them, and how such uses could be built into homework in ways that enhance learning.

**What we did:** We shared four faculty-sourced example tasks that ask students to use AI to generate (and evaluate) an outcome: generating an explanation for the special relativity barn paradox, producing and modifying Python code for creating data lists and plotting, drawing a diagram of the energy transitions in a discharge lamp, and producing a simulation of the wave functions of a particle in an infinite square well (see Figure 3).

Example 4: simulations (from PHYS 2170)

6. **[3.5pts]** Use an AI engine, preferred Chat GPT or Claude, to generate a simulation to help you visualize wave functions in an infinite square well. If using Chat GPT attempt to compile it in the Canvas.
*you are a quantum physicist seeking to make a simulation of a particle in an infinite square well. Make a simulation in html code that draws the wave function for the energy state for a given energy level. Make an input box to determine which energy level and note this with "n".*

add another feature, by following the last prompt with this one:
*now make a version where this plot of the wave function is placed inside the energy well at the appropriate energy level*

If you don't wish to use AI to make a sim, please find one on the web, or you can code one up with the standard formulas from class.

b) [1 pt] Refine the sim. Is the original correct?
If not correct, then correct it, e.g. "you should only have one energy level for each value of n" ..
If it is correct, ask it to add a feature... like time dependence, or changing the length of the well, or mass of the particle and calculate the energy levels.

c) [1 pt] Use it to check one of your answers above, e.g. problem 4. You might need to modify your sim by having it include other features. State what you did, which problem you checked, and whether or not it helped.

d) [0.5] Share your result with us. If you used Chat GPT's canvas to compile the html, hit the share button in the top right and paste the link so we can see your simulation! (or you can append to this document).

e) [0.5] what do you think of this whole exercise (just this problem not the whole problem set).

*Figure 3: an example task shared with our faculty: students use AI to generate a simulation of the wave functions of a particle in an infinite square well (PHYS 2170, Foundations of Modern Physics, sophomore level).*

In small groups, faculty worked through one or two examples of their choice with their preferred AI engine. They then reviewed provided student solutions to these tasks to identify students' prompting strategy, its strengths, and what could be improved. Groups discussed whether they found the tasks productive, whether they would adopt such AI-integrated tasks, and what guidelines and support students would need (see Figure 4).

Small group work – HW using AI generated content

4 HW examples: using AI for generating: code, diagrams, explanations, simulations.

1. Access resources through mail/QR code.
2. Choose 1-2 examples, use AI to solve.
3. Look at provided student solutions:
   - identify prompt strategy, strengths, things to improve.

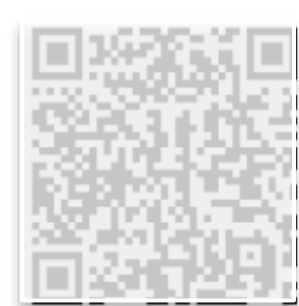

Discuss:

- Did you find this approach productive? Would you consider applying in your course?
- **What are some guidelines for using AI in HW?**
  (i.e., what do you suggest? What support do you need?)

*Figure 4: small group instructions for reflecting on HW tasks that integrate AI generated content.*

**Design principles:** the session continued discussing *approaches over tools* and enabled *local and humble inquiry* – examples were faculty-sourced and tested directly, keeping *facilitators role* in curating available materials and structuring the inquiry around them rather than disseminating best AI practices.

**What surfaced:** Working the examples raised a shared concern: students do not necessarily engage deeply with AI-generated content, and without structure they tend to shallow prompting and shallow modifications of AI generated output. The group concluded that productive use has to be modeled rather than assuming students' expertise in AI use. The session produced a set of sample approaches for using AI in coursework, added to the departmental repository.

*3.5 Session 5: AI and assessment*

Assessment was an area of great faculty concern, surfacing in almost every session. This session explicitly addressed how our assessments might adapt, in both the formative and summative approaches – we examined oral, in-person formats that make student reasoning visible, as well as an AI system for grading open-ended physics questions at scale (Kortemeyer, 2026).

**What we did:** On the formative side, a faculty member presented how he incorporated oral discussions in his course (Electricity and Magnetism 2, junior level). These were termed "mini homework defenses": short booked appointments in which a Learning Assistant (Otero et al., 2010) randomly selects one of the student's completed homework problems, the student prepares briefly and then presents and is questioned on their own solution, receiving an immediate pass/fail score (Figure 5).

Mini HW defenses (PHYS 3320)

- Purpose of the mini defenses
  - Motivate students to really understand their HW solutions (in the AI age where near-perfect solutions are readily available)
  - Give students opportunities to practice presenting physics topics (mimic ~10 min contributed talks at conferences)
- How does the mini defense work?
  - Each student books a 10-min appointment to do the mini defense (ask to see buff card)
  - The student arrives 10 min prior to prepare (can be longer with accommodations)
  - LA **randomly** choose a HW question that the student has done (verify with the student)
  - The student is allowed to use his/her own solution to the problem (digital or hardcopy)
  - The student refreshes his/her memory during the preparation time
  - The student present his/her solution during the appointment (opt-in video recording)
  - LA asks questions
  - LA gives a binary score, the student is told the score immediately
  - The student can appeal by redo the defense with the instructor
  - 85 students, 96 spots over ≥4 weeks, can cancel/rebook, or do with the instructor

*Figure 5: Some of the oral assessment guidelines ("mini-homework defences") for PHYS 3320 (Electricity and Magnetism 2, junior level), as shared by one of our faculty members.*

On the summative side, we examined an external AI-grading system for handwritten exams (Kortemeyer, 2026), used in introductory STEM courses, including its paired rubric-and-solution structure and the constraints it places on problem design and student writing, so student work can be processed reliably. We then discussed the trade-offs of using such systems, and whether and how the department might take on such platforms and what features would be most needed.

**Design principles:** By focusing on assessment we were attentive to where faculty mostly met AI *disruption* and needed support. We hold to *approaches over tools* by centering on durable assessment structures – oral defenses over particular platforms (while noting we also departed from this principle by examining an external grading platform).

**What surfaced:** Oral defenses were seen as a valuable and AI-resistant assessment option that promotes student engagement and reasoning, yet labor-intensive. Faculty were very interested in both the technical structures of the orals (timing, resources needed, graders and grading rubrics, guidelines for students) and in students' perspective on the process (which were shared through survey data). Several faculty expressed interest in and have implemented oral assessments in their future courses.

The sixth and final workshop session stepped back from specific practices to a collective reflection on the series and a discussion of where to go next as a department. We discuss these points in the next section.

## 4. Outcomes and Discussion

### *4.1 The Repository*

The series' central product is a shared departmental repository, assembled across the sessions and meant as a living resource. Currently, it holds 5 entries, each pairing curated materials with brief framing for how to use them. For example, the *AI syllabus statements* entry opens by reminding instructors why AI statements are important (regardless of their stance), and the aspects worth addressing in their statements. Then, we offer four sample policies (as requested by our faculty) spanning from no AI use to limited, conditional, and full-use, that faculty can adopt for shared language across courses, and links to other faculty policies in our department. The other entries include *Classroom discussions* – with materials to engage students in discussions around AI norms and perceptions, *Student surveys* – offering the student survey on AI use and perceptions in physics along with results across different courses, *Usage examples* – homework task with AI integration, and *Assessment* - for sharing assessment structures such as oral exams.

We shared these materials with the department at the start of the following semester to help faculty shape their AI responses, and we continue to update the repository as we curate new materials. The repository materials are further described and can be accessed through this link and upon request from authors.

### *4.2 Reflecting on the series*

We issued an end-of-series reflection survey, asking participants what aspects of the series were valuable to them, did the discussions change their thinking about AI in physics education and how, and whether they made any changes to their teaching, assessment, or course policies as a result of these discussions. Faculty most valued the collective airing of a shared problem – seeing data and approaches from different courses and surfacing concerns – more than any single response. Faculty were unanimously interested in continuing this effort. Some mentioned a shift in their stances following the series, such as recognizing they "cannot just ignore the issue" or becoming "more open to using AI for particular tasks". Fewer mentioned concrete changes as a result of the series, such as adding or refining their policy statements. As noted, following the mini-HW-defences structures shared in session 5, several faculty have taken up oral discussions in their courses the following semester, and we are now in the process of documenting and assessing this process, which will be one of the foci of this semesters' (Fall 2026) discussions.

In the collective reflection during the final session, we mapped some areas that the series have not yet addressed and we would want to address in the future, most notably the ethics of AI use and the use of AI in the instructional laboratories. Faculty also raised the question of how we could involve the colleagues who have not taken part in the series (and may be less inclined to participate), and the shared repository was in part a response to this concern.

### *4.3 An invitation*

In this work, we offer curriculum and materials that can help others start local faculty learning communities. Importantly, following our theoretical stance (Perl-Nussbaum & Finkelstein, 2026), we show that departments can begin without waiting for settled evidence on AI use, by surfacing local data, facilitating collective and humble inquiries into learning and assessment in the AI age, and treating students as partners in this effort.

We share this work not as a finished model, but as a starting point and a call for action for those interested in leading AI-related institutional change. We offer a worked example of a theoretically grounded response to AI disruption, and invite others to adapt our faculty discussion model and share back their own efforts and insights.

**Acknowledgements:**

The authors are grateful for the engagement of all of the educators involved in the learning community, for their contributions, and commitments to adapting to support our learners. This project was supported by an ISF post-doctoral fellowship (Grant No. 182/26) and a University of Colorado AI Sprint Grant.

## References:

Alba, C., Mcilwain, C., & An, R. (2026). ChatGPT on campus: how top US universities govern generative AI across higher education. *Policy Reviews in Higher Education*, 1-30.

Ben-Zion, Y., Carroll, T. K., West, C. G., Wong, J., & Finkelstein, N. D. (2026). Leveraging generative artificial intelligence for simulation-based physics experiments: A new approach to virtual learning about the real world. *Physical Review Physics Education Research*, *22*(1), 010109.

Center for Teaching & Learning. (2025). *Undergraduate perspectives on AI at CU Boulder: Report of findings.* University of Colorado Boulder. https://www.colorado.edu/center/teaching-learning/programs-services/undergraduate-student-programs/educational-technology-research-assistants-etra-0

Chasteen, S. V., Wilcox, B., Caballero, M. D., Perkins, K. K., Pollock, S. J., & Wieman, C. E. (2015). Educational transformation in upper-division physics: The Science Education Initiative model, outcomes, and lessons learned. *Physical Review Special Topics-Physics Education Research*, 11(2), 020110.

Corbin, T., Sharpe, S., & Dawson, P. (2026). On AI glasses and wearable AI in assessment. *Assessment & Evaluation in Higher Education*. https://doi.org/10.1080/02602938.2026.2661367

Corbo, J. C., Reinholz, D. L., Dancy, M. H., & Finkelstein, N. (2015). Departmental Action Teams: Empowering faculty to make sustainable change. In *2015 physics education research conference proceedings* (pp. 91-94). College Park, MD: American Association of Physics Teachers.

Cox, M. D. (2004). Introduction to faculty learning communities. *New directions for teaching and learning*, *2004*(97), 5-23.

Cruz, M., Queirós, R., Mascarenhas, D., & Ribeiro, E. (2025). Communities of Practice and Generative Artificial Intelligence in Higher Education: Pedagogical Innovation, Professional Development, and Reflective Engagement. In *International Conference on Advanced Research in Technologies, Information, Innovation and Sustainability* (pp. 279-293). Cham: Springer Nature Switzerland.

Dange, A., Lopez, R. E., Deslauriers, L., & Shah, N. (2026). aiPlato: A Novel AI Tutoring and Step-wise Feedback System for Physics Homework. *arXiv preprint arXiv:2601.09965*.

Henderson, C., Beach, A., & Finkelstein, N. (2011). Facilitating change in undergraduate STEM instructional practices: An analytic review of the literature. *Journal of research in science teaching*, *48*(8), 952-984.

Gerlich, M. (2025). AI tools in society: Impacts on cognitive offloading and the future of critical thinking. *Societies*, *15*(1), 6.

Griston, M., & Wilcox, B. R. (2025). Motivating reflection in problem solving: Homework corrections in upper-division physics courses. *Physical Review Physics Education Research*, *21*(2), 020113.

Kortemeyer, G. (2026). *Guidelines for handwritten exams with AI-grading assistance*. ETH Zurich, Rectorate and ETH AI Center. https://ethz.ch/content/dam/ethz/main/eth-zurich/education/Rector's%20Impulse%20Fund/StiftPlusGuideline.pdf

Massachusetts Institute of Technology. (2026). *Report of the Ad Hoc Committee on AI Use in Teaching, Learning, and Research Training*. https://aiandeducation.mit.edu/report/

Mazur, E. (1997). *Peer instruction: A user's manual*. Prentice-Hall.

Otero, V., Pollock, S., & Finkelstein, N. (2010). A physics department's role in preparing physics teachers: The Colorado learning assistant model. *American journal of physics*, *78*(11), 1218-1224.

Perl-Nussbaum, D., & Finkelstein, N. D. (2026). A Framework for institutional change in the age of AI. *arXiv preprint arXiv:2605.12757* [In review].

Reich, J., & Dukes, J. (2025). The future of education technology after the arrival of ChatGPT. *Phi Delta Kappan*, *107*(3-4), 19-23.

Reinholz, D. L., Pilgrim, M. E., Corbo, J. C., & Finkelstein, N. (2019). Transforming undergraduate education from the middle out with departmental action teams. *Change: the magazine of higher learning*, *51*(5), 64-70.

Skogvoll, V., & Odden, T. O. (2026). How physics professors use and frame generative AI tools. *European Journal of Physics*, *47*(2), 025713.

Schinske, J., & Tanner, K. (2014). Teaching more by grading less (or differently). *CBE—Life Sciences Education*, *13*(2), 159-166.